\documentclass[12pt,preprint]{aastex}

\shorttitle{Solar Models}
\shortauthors{Bi et al.}

\begin{document}

\title{SOLAR MODELS WITH REVISED ABUNDANCE}

\author{S. L. Bi\altaffilmark{1,2}, T. D. Li\altaffilmark{1}, L. H. Li\altaffilmark{3}, \& W. M. Yang\altaffilmark{1,4} }

\altaffiltext{1}{Department of Astronomy, Beijing Normal University,
Beijing 100875, China; bisl@bnu.edu.cn}

\altaffiltext{2}{Key Laboratory of Solar Activity, National
Astronomical Observatories, Chinese Academy of Sciences}

\altaffiltext{3}{Department of Astronomy, Yale University, P.O. Box
208101, New Haven, CT 06520-8101, USA}

\altaffiltext{4}{School of Physics and Chemistry, Henan Polytechnic
University, Jiaozuo 454000, Henan, China}

\begin{abstract}

We present new solar models in which we use the latest low
abundances and we further include the effects of rotation, magnetic
fields and extra-mixing processes. We assume that the extra-element
mixing can be treated as a diffusion process, with the diffusion
coefficient depending mainly on the solar internal configuration of
rotation and magnetic fields. We find that such models can well
reproduce the observed solar rotation profile in the radiative
region. Furthermore the proposed models can match the seismic
constraints better than the standard solar models, also when these
include the latest abundances, but neglect the effects of rotation
and magnetic fields.

\end{abstract}


\section{Introduction}

The standard solar models with the latest input physics are well
known to yield the solar structure to an amazing degree of
precision, and agree with the helioseismic inversions (see e.g.,
Christensen-Dalsgaard et al. 1996; Bahcall et al. 2001). Those
models use the old abundance values (Grevesse \& Savual 1998,
hereafter GS98). However, the standard solar models with the new
solar mixture AGS05 (Asplund et al. 2005, hereafter AGS05) disagree
with helioseismic constraints, e.g., the position of the base of
convection zone (CZ) is too shallow and the surface helium abundance
is lower than in the Sun \citep{chr91,bas98,bas04}. Larger
discrepancies are in the sound-speed and density profiles between
the Sun and the models with low $Z$ \citep{bah04,guz05}. See
\citet{bas08} for a detailed review paper.

A number of investigations have attempted to explain this
discrepancy as a matter of improved physical inputs in the standard
solar model such as: enhanced diffusivity, opacity increases,
convective overshooting, low-Z accretion
\citep{bah05,bahc06,yan07,chr09,ser09,guz10,tur10} and revised solar
composition (Asplund et al. 2009, hereafter AGSS09). The results of
all of these analyses show that it is difficult reproduce the
helioseismic constraints with the standard solar model also when it
includes the latest abundances. This failure could be due to the
fact that the standard solar model neglects rotation, magnetic
fields and some extra-mixing processes. In this letter, we show that
these effects can reduce the discrepancies.

\section {Beyond the standard solar model}

Helioseismology has revealed that the Sun is rotating differentially
at the surface, slowly in the core and almost uniformly in the
radiative region \citep{cha99}. The purely rotation-induced mixing
has been considered in modeling rotating stars, as given in
\citet{zah92}, \citet{mae98}, \citet{mae00}, \citet{pal03} and
\citet{mat04}. However, these models appear insufficient to
reproduce the helioseismically inferred internal solar rotation
profile. This suggests that other effects should be considered in
extracting angular momentum from the central core of the Sun.

Recently, two main mechanisms have been proposed to explain the
solar flat rotation profile, namely internal gravity waves (e.g.
Charbonnel \& Talon 2005) and magnetic fields (e.g. Eggenberger et
al. 2005). Here, we mainly describe the efficiency of the
extra-mixing caused by rotation and magnetic fields, as prescribed
by the Tayler-Spruit dynamo \citep{pit86,spr02}. The theoretical
formulation of this dynamo is still a matter of debate
\citep{den07,zah07}, however, \citet{egg05} found that the model
with the Tayler-Spruit dynamo-type field successfully reproduces the
observed solar rotation profile. Therefore, it is particularly
interesting to investigate the effects of rotation and magnetic
fields on the solar interior and global parameters.

We present a simple scheme for dealing with angular momentum
transport and element mixing in the solar interior. It is based on
the stellar structure equations which include rotation and magnetic
fields \citep{pin89,li03}. The detailed derivation is given in
\citet{yan06,yan08}. This formulation allows us to estimate the
effects of rotation and magnetic fields on the Sun properties. The
angular momentum transport and elements mixing can be described with
two diffusion equations as follows:
\begin{eqnarray}
\rho r^{2}\frac{\partial \Omega}{\partial t} &=& f_{\Omega}
\frac{1}{r^{2}}\frac{\partial}{\partial r}\left[\rho
r^{4}\left(D_{rot}+D_{m}\right)\frac{\partial \Omega}{\partial
r}\right], \\
\frac{\partial X_{i}}{\partial t}&=& f_{C}\frac{1}{\rho
r^{2}}\frac{\partial}{\partial r}\left[\rho
r^{2}(D_{rot}+D_{m}^{'})\frac{\partial X_{i}}{\partial
r}\right]+\left(\frac{\partial X_{i}}{\partial
t}\right)_{nuc}+\left(\frac{\partial X_{i}}{\partial
t}\right)_{micro},
\end{eqnarray}
where the adjustable parameters $f_{\Omega}$ and $f_{C}$ are
introduced to represent some inherent uncertainties in the diffusion
equations. The second and third terms on the right-hand-side of
Equation (2) are the nuclear and gravitational settling terms,
respectively. In our model, the diffusion coefficient $D_{rot}$ is
associated with the rotational instability as described by
\citet{cha95}. In the case of a Tayler-Spruit dynamo-type field, the
diffusion coefficient for the angular momentum transport can be
written as \citep{mae03}:
\begin{eqnarray}
D_{m}&=& r^{2}\Omega q^{2}\left(\frac{\Omega}{N_{\mu}}\right)^{4}
\end{eqnarray}
and the one for chemical element transport as:
\begin{eqnarray}
D_{m}^{'}&=& r^{2}\Omega
q^{4}\left(\frac{\Omega}{N_{\mu}}\right)^{6}.
\end{eqnarray}
Equations (3) and (4) are valid in the regime of negligible thermal
diffusion, namely $N_{\mu}\gg N_{T}$, where $N_{T}$ ($N_{\mu}$)
represents the thermal ($\mu$-) gradients associated with buoyancy
frequency. When this condition is violated, we should replace
Equations (3) and (4) with
\begin{eqnarray}
D_{m}&=&r^{2}\Omega\left(\frac{\Omega}{N_{T}}\right)^{1/2}
\left(\frac{K}{r^{2}N_{T}}\right)^{1/2}
\end{eqnarray}
and
\begin{eqnarray}
D_{m}^{'}&=&r^{2}\Omega\left|q\right|\left(\frac{\Omega}{N_{T}}\right)^{3/4}
\left(\frac{K}{r^{2}N_{T}}\right)^{3/4},
\end{eqnarray}
respectively, where $q=-\frac{\partial \ln\Omega}{\partial\ln r}$
and $K=4acT^{3}/3\kappa\rho^{2}c_{p}$ is the thermal diffusivity.

Additionally, in order to reproduce the solar surface angular
velocity, we adopt the \citet{kaw88} braking law:
\begin{equation}
\frac{dJ}{dt}=f_{K}K_{\Omega}\left(\frac{R}{R_{\odot}}\right)^{1/2}
\left(\frac{M}{M_{\odot}}\right)^{-1/2}\Omega^{3},
\end{equation}
where $K_{\Omega}\simeq 1.13\times 10^{47}$ g cm$^{2}$ s and $f_{K}$
is an adjustable parameter related to the magnitude of the magnetic
fields.

\section {Calculations and results}

Our solar models are obtained from the one-dimensional Yale Rotating
Stellar Evolution Code (YREC; Guenther et al. 1992; Li et al. 2003;
Yang \& Bi 2006), by including rotation, magnetic fields and
relevant extra-mixing processes. In addition, we use the following
updated physical quantities: OPAL equation of state tables EOS2005
\citep{rog02}, the opacities (GS98, AGS05 and AGSS09) supplemented
by the low-temperature opacities \citep{fer05}, diffusive element
settling \citep{tho94} and the Krishna-Swamy Atmosphere $T-\tau$
relation.

In order to investigate the influence of rotation and magnetic
fields, we constructed these solar models in accord with different
physical processes corresponding to different solar compositions. In
the numerical calculations, all models are calibrated from the
initial zero-age main sequence (ZAMS) to the present solar-age
models, for which the radius is $6.9898\times10^{10}$ cm, the
luminosity $3.8515\times10^{33}$ erg/g, the mass
$1.9891\times10^{33}$ g and the adopted photospheric $Z/X$ ratio.
The free variables are the initial helium abundance $Y$, the initial
metallicity $Z$ and the mixing-length parameter, all of which are
adjusted to match these observational constraints. In addition, we
assumed that the convective region rotates rigidly, as proposed by
\citet{pin89}. The initial angular velocity is another free
parameter that can be tuned so that the surface velocities of
solar-age models match the observed values.

Figure 1 shows the angular velocity as a function of radius $r$ at
the ages of 2.0 Gyr and 4.57 Gyr. For the purely rotating model, the
$\Omega$-gradient clearly appears in the radiative region. In the
solar interior, the angular velocity increases with increasing age
during the main sequence, while in the surface it is just the
opposite. As a consequence of this at the present age the core
rotation velocity is about four times as large as the surface one.
On the other hand, the angular velocity profile for the model with
magnetic fields is significantly different. During the main-sequence
stage, the Sun is a quasi-solid body. It is interesting to note that
at the age of 4.57 Gyr, the surface rotation velocity predicated by
both models is approximately $2.9 \times 10^{-6}$ rad/s. However,
the total angular momentum is quite different. For the calibrated
models, the total angular momentum of the rotating model at the age
4.57 Gyr is $8.97 \times 10^{48}$ g cm$^{2}$ s$^{-1}$, which is
about five times as large as the seismic result $(1.94 \pm
0.05)\times 10^{48}$ g cm$^{2}$ s$^{-1}$ \citep{kom03}; while for
the model with magnetic fields at the same age, the total angular
momentum is $2.02 \times 10^{48}$ g cm$^{2}$ s$^{-1}$, which is in
good agreement with the result obtained by helioseismology at
$1\sigma$ level. The magnetic field thus constitutes a more
efficient process to transport angular momentum because it enhances
the coupling between the radiative zone and convective one.

Rotation and magnetic fields have important consequences on the
chemical composition profile of the outer convective envelope, as
shown in Figure 2. By comparing the models with and without rotation
and magnetic fields, we find that the extra-mixing process
counteracts the effect of diffusive settling in the outer envelope.
Hence, the model with rotation and magnetic fields has a smoother
helium abundance profile than the standard solar model with the same
abundance. This leads to a change in the CZ structure, as well as
improvements in the sound speed and density profiles.

For further investigation of the magnetic field's role, Figure 3
shows in detail the differences between the calculated and inferred
sound-speed and density profile \citep{bas09}. The different lines
refer to calibrated evolved models at the age of 4.57 Gyr, each
using a different abundance, indicated as GS98, AGS05 and AGSS09. It
is clearly visible that the discrepancy between seismic inferences
and solar models with the new lower abundances is much larger than
one with the old abundance. As illustrated by the different curves,
the effects of rotation and magnetic fields on the stellar structure
equations change the hydrostatic equilibrium and thermodynamic
variables on the solar interior, and therefore also have a
significant impact on the solar models. Table 1 summarizes the main
characteristics of our calibrated models. Interestingly, among all
the models listed in the table, model AGSS09c shows the best
agreement with the inversions. This model reproduces the sound speed
and density profiles in within $0.5\%$, while for model GS98a the
discrepancy is about $0.3\%$. Furthermore, model AGSS09c predicts
the position of the CZ base at $R_{CZ}=0.721R_{\odot}$ which shows a
$8 \sigma$ discrepancy, while AGSS09a model shows a $10 \sigma$ one.
For the surface helium abundance the situation is analogous: model
AGSS09c predicts $Y_{s}=0.243$ with a $1.5 \sigma$ discrepancy,
while model AGSS09a shows the discrepancy at $3.6 \sigma$ level.
Although the models including rotation and magnetic fields show some
improvements with respect to the standard solar model, they still
disagree with the seismic constraints.

\section {Conclusions}

We have investigated the effects of rotation and magnetic fields on
the solar models, and have found that when these effects are
included, alongside the new abundances, the revised solar models can
better reproduce the helioseismic constraints. However, we see that
it is difficult to match simultaneously the new abundances and
helioseismology data for sound speed, density profiles, convection
zone depth and surface helium abundance. Although the Tayler-Spruit
dynamo type magnetic field still needs to be studied further, our
results show that it does provide a possible explanation for the
solar abundance problem. We have neglected turbulence, which may
feed the differential rotation and sustain magnetic fields in the
convection zone, and other interactions. These physical processes
will be considered in our future work. The results obtained in this
paper are encouraging and we intend to apply our model to solar-type
stars to get a proper interpretation of the existing helioseismic
observations and the coming asteroseismic ones.

\acknowledgments S.L.B. acknowledges grant 2007CB815406 of the
Ministry of Science and Technology of the Peoples Republic of China
and grants 10773003 and 10933002 from the National Natural Science
Science Foundation of China. L.H.L. acknowledges the financial
support of Grant ATM 073770 by NSF and the Vetlesen Foundation of
USA.

\begin{deluxetable}{cccccccccccccc}
\tablecolumns{13} \tabletypesize{\scriptsize}
 \tablewidth{0pt}
\tablecaption{Characteristics of the Calibrated Solar Models}
\tablehead{\colhead{Model} & \colhead{$(Z/X)_{s}$} &
\colhead{${Z_s}$} & \colhead{$Y_{s}$} & \colhead{$R_{cz}/R_{\sun}$}
& \colhead{$<\delta c/c>$} & \colhead{$<\delta \rho/\rho>$} &
\colhead{$Y_{c}$} & \colhead{$Z_{c}$} & \colhead{$Y_{ini}$} &
\colhead{$Z_{ini}$} & \colhead{$\alpha_{MLT}$}} \startdata
  GS98a  & 0.0229 & 0.0169 & 0.246 & 0.715 & 0.0012 & 0.008 & 0.644 & 0.0198 & 0.277 & 0.0188 & 2.12 \\
  AGS05a & 0.0165 & 0.0125 & 0.230 & 0.728 & 0.0030 & 0.034 & 0.623 & 0.0148 & 0.261 & 0.0140 & 2.08 \\
  AGS05b & 0.0165 & 0.0124 & 0.239 & 0.727 & 0.0028 & 0.035 & 0.622 & 0.0146 & 0.269 & 0.0139 & 2.04 \\
  AGS05c & 0.0165 & 0.0124 & 0.237 & 0.726 & 0.0028 & 0.033 & 0.621 & 0.0146 & 0.260 & 0.0139 & 2.05 \\
  AGSS09a & 0.0181 & 0.0136 & 0.236 & 0.723 & 0.0020 & 0.022 & 0.631 & 0.0160 & 0.267 & 0.0152 & 2.12 \\
  AGSS09b & 0.0181 & 0.0134 & 0.245 & 0.722 & 0.0019 & 0.023 & 0.630 & 0.0158 & 0.268 & 0.0150 & 2.07 \\
  AGSS09c & 0.0181 & 0.0135 & 0.243 & 0.721 & 0.0017 & 0.021 & 0.630 & 0.0158 & 0.266 & 0.0150 & 2.09 \\
 \enddata
\tablenotetext{a}{Solar models with diffusion.}
\tablenotetext{b}{Solar models with diffusion and rotation.}
\tablenotetext{c}{Solar models with diffusion, rotation and magnetic
fields.}
\end{deluxetable}

\clearpage

\begin{figure}
\epsscale{0.60} \plotone{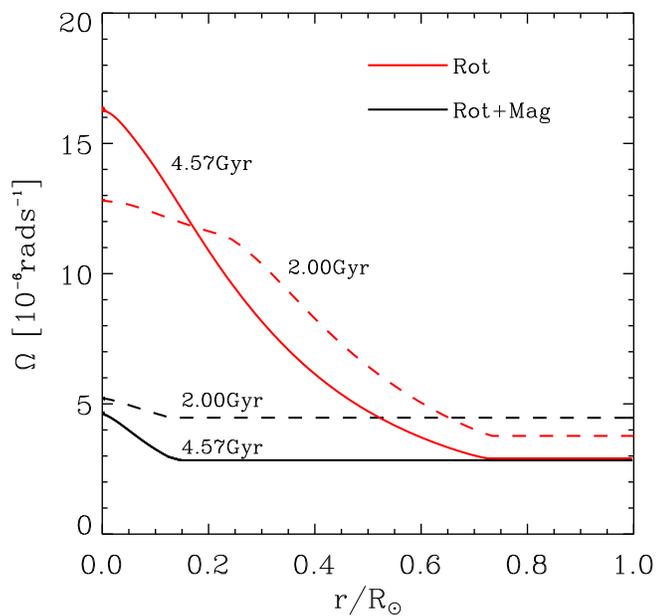} \caption{Comparison of angular
velocity profiles for the two cases: (1) purely rotating models; and
(2) models with rotation and the Tayler-Spruit dynamo-type field.
The dashed and solid lines refer to different ages: 2.0 Gyr and 4.57
Gyr, respectively. \label{fig1}}
\end{figure}

\clearpage

\begin{figure}
\epsscale{0.60} \plotone{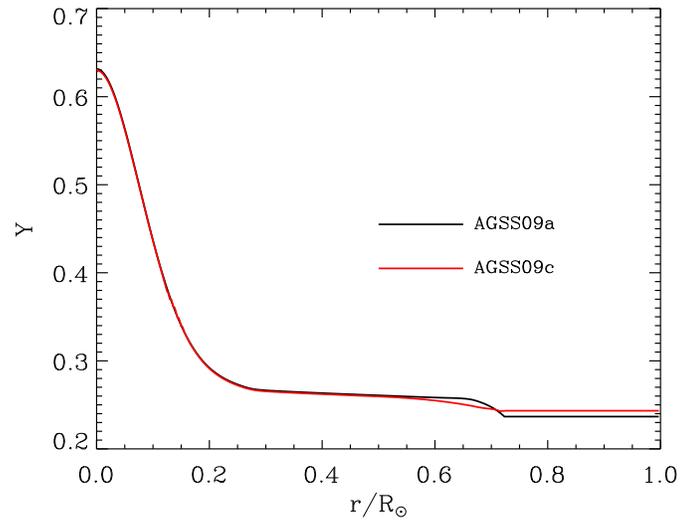} \caption{Helium abundance
profiles for the calibrated solar models at the age of 4.57 Gyr,
computed with and without rotation and magnetic fields.
\label{fig2}}
\end{figure}

\clearpage

\begin{figure}
\epsscale{0.60} \plotone{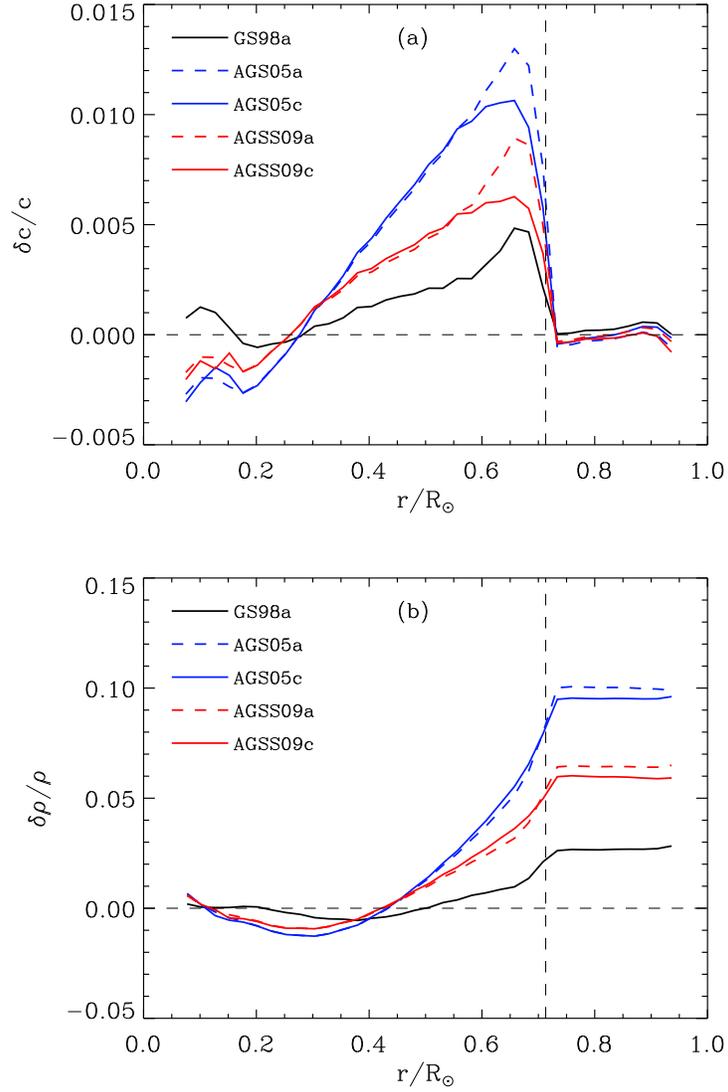} \caption{Differences between
inferred and calculated sound speeds and densities for models with
and without rotation and magnetic fields at the age of 4.57 Gyr,
corresponding to the GS98, AGS05 and AGSS09 abundances. Sound speed
and density inversions are from Basu et al. (2009).\label{fig3}}
\end{figure}

\end{document}